\documentclass[12pt]{article} 
\usepackage{amssymb}
\usepackage{amsmath}
\usepackage{eucal}
\usepackage{epsfig}

\def\be{\begin{equation}}
\def\eeq{\end{equation}}
\def\bea{\begin{eqnarray}}
\def\eea{\end{eqnarray}}

\def\re#1{(\ref{#1})}

\begin{document}

\begin{titlepage}

\rightline{hep-th/0607246}
\vskip 1. cm
\vskip 1 cm
\begin{center}
\huge{\bf{Elementary Kaluza-Klein Towers revisited   }}
\end{center}
\vskip 0.6 cm
\begin{center} {\Large{Fernand Grard$^1$}},\ \ {\Large{Jean Nuyts$^2$}}
\end{center}
\vskip 2 cm

\noindent{\bf Abstract}
\vskip 0.2 cm
{\small 
\noindent Considering that the momentum squared in the extra dimensions is
the physically relevant quantity for the generation of the Kaluza-Klein
mass states, we have reanalyzed mathematically
the procedure for five dimensional scalar fields within the Arkhani-Ahmed,
Dimopoulos and Dvali scenario. 
We find
new sets of physically allowed boundary conditions. 
Beside the usual results, they lead to new towers 
with non regular mass spacing, to lonely mass states and to tachyons. We remark that, 
since the SO(1,4) symmetry 
is to be broken due to the compactification of the extra dimensions, 
the speed of light could be
different in the fifth dimension. This would lead to the possible appearance of a new universal 
constant besides $\hbar$ and $c$. 
 }
\vfill
\noindent
{\it $^1$  Fernand.Grard@umh.ac.uk,  Physique G\'en\'erale et Physique des Particules El\'ementaires,
Universit\'e de Mons-Hainaut, 20 Place du Parc, 7000 Mons, Belgium}
\vskip 0.2 cm
\noindent
{\it $^2$  Jean.Nuyts@umh.ac.be, Physique Th\'eorique et Math\'ematique,
Universit\'e de Mons-Hainaut, 20 Place du Parc, 7000 Mons, Belgium}
\end{titlepage}

\section{Introduction}{\label{Intro}}

An extra-dimension scenario has been proposed by Arkhani-Ahmed,
Dimopoulos and Dvali
\cite{ADD}
to try to solve the mass scale hierarchy problem: the Planck mass
scale
$M_{Pl}\equiv\sqrt{\hbar c/G_N}= 1.22090(9)\  10^{19}$GeV
($\overline{M_{Pl}}=M_{Pl}/\sqrt{8\pi}$)
is considerably larger
than the weak mass scale $M_{EW}\approx 1$Tev.

In this scenario, it is usually
assumed that the full space is a $(4{+}n)$-dimensional space which is
taken to be flat and real. The standard model particles are
constrained to propagate in our four dimensional world (called a
``wall''
or a ``brane'') whereas gravity only is allowed to propagate in the
full
$(4{+}n)$-dimensional ``bulk''.
It is then argued that the Planck mass would be related to the
$(4{+}n)$-dimensional fundamental scale $M_{*}$ by the relation
\be
\overline{M_{Pl}}^2\approx V_n M_{*}^{n+2}\ ,
     \label{mpla}
\end{equation}
where $V_n$ is the $n$-dimensional volume of the $n$ extra dimensions.
Assuming equal size $2 \pi R$ for each of them, then
\be
V_n=(2\pi R)^n\ .
     \label{mplb}
\end{equation}
The speed of light $c$ and $\hbar$ are used as the only two basic
constants of the theory and
$M_{*}$ is chosen of the order of $M_{EW}$. It is
usually concluded that $n=1$ is excluded. The size $2\pi R$ of the
extradimension would then be of the order of $10^8$m and much too
large as
Newton's law is known to hold very well for the sun planets.
For $n=2$, the order of magnitude of $2\pi R$ becomes much smaller,
about $100 \mu$m, close to
where Newton's law has been tested \cite{HKH} with a sufficient
precision.
It has been argued that the ADD scenario provides experimental tests.
They come through the prediction of the so called Kaluza-Klein mass
towers related
to the massless graviton.

Similarly an alternative scenario has been proposed by Randall and Sundrum \cite{RS} in which 
the mass scale hierarchy results from an exponential warp factor 
in a five dimensional non factorizable geometry. The idea that the Standard Model gauge 
and matter fields propagate in the bulk has also been considered \cite{DHR}. 

Experimental searches for extra dimensions have been carried on at
the CERN LEP \cite{CERN}, at the TEVATRON \cite{TEVATRON} and at HERA \cite{HERA}. 
They conclude that no statistically significant extra dimension effects have been
observed. Further investigations are expected to be persued at the LHC \cite{LHC}.

Here we follow the ADD scenario in the
five dimensional case ($n=1$). We call $s$ the extra dimension
supposed to be space like in order to have only one time.
Indeed, considerations of two times lead to many difficulties,
the ensuing possible breaking of causality being an important one
\cite{DGS}.
We restrict ourselves
to scalar particles,
originating from 
a real five dimensional massless scalar field assumed to propagate in the bulk. 
The associated Kaluza-Klein 
masses $m_n$ are deduced from a natural ans\"atz expressed by the equation
\be
\partial^2_s \phi^{[n]}n=-m_n^2 \phi^{[n]}\ .
\label{meq}
\end{equation}
The mass eigenvalues $m_n^2$ are essentially determined
by boundary conditions imposed on the scalar fields.
As summarized for example by Rizzo \cite{R},
periodic boundary conditions
where the scalar field moves around a circle in $s$ 
of radius $2\pi R$,
box type conditions where the scalar particle is confined to a strip
of $s$
of finite size or other boundary conditions, can be imposed.
They lead to the specific physical spacing of the masses
of the fields present in the tower.
In particular, they govern the presence or absence
of a zero mass field at the bottom of the tower.

In this article, starting from basic physical considerations,
we study the minimal set of coherent physical requirements
which should lead to the boundary conditions.
We transpose these physical ideas in a precise mathematical
sense and discuss all the ensuing allowed boundary conditions
and the resulting physical Kaluza-Klein states.

The article is organised as follows.
In section \re{bounda}, we present and discuss the physical
ideas.
From this discussion, all the allowed boundary conditions are deduced.
In section \re{KKm}, we construct
the scalar mass states which follow from the boundary conditions applied
to a real massless scalar field in five dimensions. 
Some towers are new \re{towers} and lonely KK states are expected \re{Lonely}. 
For completeness, we show that tachyon
states may exist \re{tachyons} for certain specific boundary conditions.
In section \re{examp}, we show explicitly on one example
how one of the new towers develops.
Finally, in section \re{conclu}, we stress our new findings
and outline the interest of the extension of our approach 
to particles of higher spin.

\section{Physical considerations and boundary conditions}
{\label{bounda}}

As stated above, we
limit ourselves to a discussion involving one extra dimension
labeled by $s$ and supposed to be space like.
It is assumed that, though in our space the variables
$x^{\mu}$ ($\mu=0,\dots,3$) appear to span the full $[-\infty,+\infty]$ range, the
extra dimension $s$ is of finite extension $[0,2\pi R]$.
It should be remarked immediately that this hypothesis
destroys the generalized
Minkowski SO$(1,4)$ invariance which could be potentially present.
The minimal symmetry is therefore restricted to SO(1,3).

Essentially one can then imagine three major types of geometries.
The space is like a ribon and our brane is at one edge
in the extra dimension,
say at the point $s=0$.
The space is like a cylinder and the point $\{s=0\}$ is
identical to the point
$\{s=2\pi R\}$. Our brane is located say at that point $s=0=2 \pi R$
and is a
generator of the cylinder. The space is like a ribon and our brane is not at an
edge but somewhere inside, say at the point located at some point
$\hat{s},\ 0<\hat{s}<2\pi R$ within the allowed $s$ range.

Though the SO(1,4) symmetry is broken from the start,
it is usually assumed that $s$ and $x^i$ ($i=1,2,3$) are commensurable.
To study in a precise way a minimal deviation from this symmetry,
let us first define the length squared in the five dimensional space as 
by
\be
L^2=(ct)^2-\sum_{i=1}^{3}(x^i)^2-a^2 s^2\ .
\label{length}
\end{equation}
It is obvious, using the variables to $t'=ct,\ {x'}^i=x^i,\ s'=as$, that
this distance is fully SO(1,4) invariant. The natural invariant equation for a 
massless scalar field is then 
\be
\left(\frac{1}{c^2}\partial_{t}^2 - \sum_{i=1}^{3}\partial_{x_i}^2 -
      \frac{1}{a^2}\partial_s^2\right)\Phi(x_{\mu},s)=0
\label{massless2}
\end{equation}
and the speed of light along the $s$ direction is equal to the speed of light in the $x^i$ directions.

A minimal deviation from the SO(1,4) symmetry would be to replace the parameter $a$ in \re{massless2} by
an independent parameter $b$ in the form
\be
\left(\frac{1}{c^2}\partial_{t}^2 - \sum_{i=1}^{3}\partial_{x_i}^2 -
      \frac{1}{b^2}\partial_s^2\right)\Phi(x_{\mu},s)=0\ .
\label{massless3}
\end{equation}
The speed $c$ of a massless particle in the $x^i$ direction is different
from the speed $ca/b$ in the $s$ direction. The parameter $b/a$,
a priori unknown and unpredictable, is a measure 
of the violation of SO(1,4) and of the commensurability of the length 
units in the $x^i$ and $s$ directions. 

To keep things as simple as possible, we have put $a=b=c=1$.
The parameter $a/b$ does not play any role in the mathematics 
that follows.
The possibility $a/b\neq 1$ should be kept in mind
once experimental consequences are looked for and masses are evaluated 
in terms of $1/R$.

We then follows the usual pattern,
supposing a separation of variables
in the form (where $n$ is some summation index)
\be
\Phi(x_{\mu},s)=\sum_n\psi^{[n]}(x_{\mu})\phi^{[n]}(s)
\label{varseparation}
\end{equation}
with $\psi^{[n]}(x_{\mu})$ and $\phi^{[n]}(s)$
both real by construction.

The key ans\"atz of the approach is to suppose that
$\phi^{[n]}(s)$ satisfies the equation
\be
\partial_s^2\phi^{[n]}(s)=-m_n^2\phi^{[n]}(s)
\label{phisequation}
\end{equation}
with $m_n^2$ real. 
One then finds from Eq. \re{massless2} (with $a=1$)
\be
(\square_{4}+m_n^2)\psi^{[n]}(x_{\mu})=0\ .
\label{phixequation}
\end{equation}
If $m_n^2$ is assumed to be positive (or zero),
this corresponds to an ordinary 
massive (or massless) scalar four dimensional particle,
a so-called Kaluza-Klein mass state.
If $m_n^2$ happens to be negative, this would correspond to a tachyon.

On the finite $s$ region, the Eq. \re{phisequation} is usually solved
by
imposing,
for example, periodic, antiperiodic or box like boundary conditions
\bea
\left\{\begin{array}{rcl}
        \phi^{[n]}(2\pi R)&=&\phi^{[n]}(0)\\
        \partial_s\phi^{[n]}(2\pi R)&=&\partial_s\phi^{[n]}(0)
       \end{array}\right\}&\quad&
              {\rm{Periodic\ Boundary\ Conditions}}\ ,
    \nonumber\\
\left\{\begin{array}{rcl}
        \phi^{[n]}(2\pi R)&=&-\phi^{[n]}(0)\\
        \partial_s\phi^{[n]}(2\pi R)&=&-\partial_s\phi^{[n]}(0)
       \end{array}\right\}&\quad&
              {\rm{Antiperiodic\ Boundary\ Conditions}}\ ,
    \nonumber\\
\left\{\begin{array}{rcl}
        \phi^{[n]}(0)&=&0\\
        \phi^{[n]}(2\pi R)&=&0
       \end{array}\right\}&\quad&
              {\rm{Box\ Conditions}}\ .
\label{Pself}
\eea
All these conditions follow essentially from the requirement
that the operator $P^s=i\partial_s$ as well as its square $P_s^2$
are symmetric as it is usually assumed 
for the energy operator and momentum operators $P^{\mu}$
in our four dimensional space.

One may wonder if this is really the most general physical
requirement.
Indeed, in view of the fact that
the physically measurable quantities in
our four dimensional subspace are the
$m_n^2$ arising in Eq. \re{phixequation},
which hence have to be real, we conclude that the minimal requirement
is to have $P_s^2$ symmetric and not necessarily $P_s$ itself.

It should be stressed that if on an infinite dimensional real space
($[-\infty,\infty]$ range)
the notion of symmetric and selfadjoint momentum operators are essentially
identical,
this is not automatically the case on a space with
finite extent \cite{RN}.

Our physical hypothesis translates then into the mathematically
more precise hypothesis
\be
\left\{
{\rm{Basic\ Hypothesis}}:\quad P_s^2 {\rm{\ is\ a\ symmetric\
operator}}
\right\}\ .
\label{bashyp}
\end{equation}

Let $\phi_a(s)$ and $\phi_b(s)$ be two vectors
in the relevant Hilbert space with scalar product
(remember that we have restricted ourselves to real fields)
\be
(\phi_a,\phi_b)\equiv \int_{0}^{2\pi R} \phi_a(s)\phi_b(s)\ ds\ .
\label{scalprod}
\end{equation}
The basic hypothesis \re{bashyp} is equivalent to the statement
\be
(\phi_a,P_s^2 \phi_b)=(P_s^2 \phi_a,\phi_b)
\label{P2adj}
\end{equation}
which leads to the conditions
\bea
&
\Bigl[\left(\partial_s\phi_a\right)\phi_b \Bigr]_{s=2\pi R}
   -
\Bigl[\left(\partial_s\phi_a\right)\phi_b\Bigr]_{s=0}
&
       \nonumber\\
&&     \nonumber\\
&\quad\quad \ =
         \Bigl[  \phi_a\left(\partial_s\phi_b\right)\Bigr]_{s=2\pi R}
    -
\Bigl[ \phi_a\left(\partial_s\phi_b \right)\Bigr] _{s=0}& \ .
\label{selfadjsq3}
\eea

From this equation, all the specific boundary conditions which
guarantee that $P_s^2$
is symmetric can easily be deduced. Boundary conditions must relate
linearly
and homogeneously the four quantities (notation $\partial_s\phi(a)\equiv \partial_s\phi|_{s=a}$)
\be
    \phi(0),\quad \phi(2\pi R), \quad \partial_s\phi(0),\quad
    \partial_s\phi(2\pi R)\ .
\label{quantities}
\end{equation}
In the Hilbert space of square integrable functions
on $[0,2\pi R]$, they constrain the vectors to be in the domain of a
symmetric $P_s^2$ operator.
Hence the same relations have to hold for $\phi_a$ and $\phi_b$
independently.

It is convenient to build all possible sets of boundary conditions
and classify them canonically according to
the number of independent homogeneous linear relations
they are supposed to satisfy.
The general discussion leads to numerous cases with
four relations (Case indexed by A), with
three relations (Cases indexed by B) and with
two relations (Cases indexed by C).
Eq. \re{selfadjsq3} cannot be satisfied if there is one relation
only.

The results are summarized in Table \re{tablebound}.
The
$\lambda_i$ ($i=1,2,3$), $\mu_i$ ($i=1,2$),
$\alpha_i$ ($i=1,\dots,4$),
$\rho_i$ ($i=1,2$),
$\nu$, $\kappa$ and $\zeta$ are arbitrary
real parameters which specify the relevant domains
of the corresponding operators $P_s^2$.

\section{Kaluza-Klein mass states}{\label{KKm}}

In this section we consider and solve the basic equation \re{phisequation}
which determines the masses of the
Kaluza-Klein states. There are mass towers, lonely states and tachyons.
Remind that all the boundary conditions of
Table \re{tablebound} derive from the imposition that the operator
$P_s^2$ is symmetric and hence that its eigenvalues $m_n^2$ are real.

\subsection{Towers}{\label{towers}}

Let us first suppose that $m_n^2$ is positive corresponding 
to ordinary scalar particles. 
Eq. \re{phisequation} is trivial to solve irrespective of
the boundary conditions
\be
\phi_n(x^{\mu},s)=\sigma_n\,\sin(m_ns)
   +\tau_n\,\cos(m_ns)
\label{soleqnew1}
\end{equation}
where $\sigma_n$ and $\tau_n$ are real functions of $x^{\mu}$.
In fact, for the same mass squared value there exists two independent
eigenvectors where the variables separate. The derivative writes trivially
\be
\partial_s\phi_n(s)=m_n\sigma_n\,\cos(m_ns)-m_n\tau_n\,\sin(m_ns)\ .
\label{soleqnew2}
\end{equation}

We introduce the notation
\bea
S_n&=&\sin(2\pi m_n R)
    \nonumber\\
C_n&=&\cos(2\pi m_n R)
    \nonumber\\
S_n^2+C_n^2&=&1
\label{notation1}
\eea
and, dropping the index $n$ for the time being, write
\bea
\phi(0)&=&\tau
    \label{notation2a}\\
\partial_s\phi(0)&=&m\sigma
    \label{notation2b}\\
\phi(2\pi R)& =&\sigma S+\tau C
    \label{notation2c}\\
\partial_s\phi(2\pi R)&=&m\sigma C-m\tau S\ .
    \label{notation2d}
\eea

It is then a matter of simple algebra to replace these quantities
successively in the different allowed boundary relations enumerated in
Table \re{tablebound}. Each of the cases obviously leads to
restrictions for the values of the functions $\tau$ and $\sigma$ and
the parameter $m$
of the solutions and relates them to the size $2\pi R$ 
of the $s$-space. In a few cases, the boundary conditions are inconsistent with
the form of the solution. In other cases, the compatibility of the
equations restricts the parameters of the allowed boundary conditions.
In particular, four boundary relations (Case A of
Table \re{tablebound}) do not lead to any non trivial solution.
Many types of situations occur.
In general, an infinite set of masses constituting the Kaluza-Klein
towers appears. For boundary conditions involving three relations they
are summarized in Table \re{tablethree} and for two relations in
Table \re{tabletwo}. In almost all of these cases there is no field of
zero mass at the bottom of the tower. It should be remarked that
in the classical
Kaluza-Klein towers, i.e. when $P_s$ is selfadjoint,
the spacing between the masses is regular. Here, as will be seen in an
example in section \re{examp}, in many cases deviations from the
regular spacing occur.
In a small set of cases, the infinite tower has a mass zero state at
its bottom. We have grouped these cases together in
Table \re{tablezero}.

\subsection{Lonely masses} {\label{Lonely}}

In a few cases, a lonely state appears when very specific relations are satisfied between
the parameters of the boundary conditions and the size $2\pi R$ of the $s$ region.
These lonely states are summarized in Table \re{tablespecial}.

\subsection{Tachyons}{\label{tachyons}}

In the preceding subsections we have solved
the basic equation \re{phisequation}
supposing that the mass squared eigenvalues of the symmetric operator of
$P_s^2$ are positive or zero. We here study the case of $m_n^2$
negative and write $h_n^2=-m_n^2$. The solutions, when they exist,
will correspond to tachyons arising for certain types of boundary
conditions.
Eq. \re{phisequation} is again trivial to solve irrespective of
the boundary conditions. The real solutions (dropping the $n$ index
for simplicity) are
\be
\phi(s)=\widetilde{\sigma}\,e^{hs}
   +\widetilde{\tau}\,e^{-hs}
     \label{soleqtach1}
\end{equation}
where $\widetilde{\sigma}$ and $\widetilde{\tau}$ are real.

Going through all the sets of boundary conditions, one finds in
a certain number of cases lonely tachyons where the $h^2$ as well as
the size $2\pi R$ of the $s$ region are fixed in terms of the
parameters. They are summarized in Table \re{tabletachyons1} for the
cases corresponding to three boundary relations, and in
Table \re{tabletachyons2} for two boundary
relations.

\subsection{Example}{\label{examp}}

As an explicit example of the mass structure of the new Kaluza-Klein
towers,
we consider the Case B1b-tw (see Table \re{tablethree}). The masses $m_n$ are the solutions of
the equation
\bea
\tan(2\pi m_n R)&=&\frac{2 m_n\lambda}{m_n^2-\lambda^2}
    \nonumber\\
{\rm{sign}}(\sin(2\pi m_n R))&=&\epsilon\ {\rm{sign}}(\lambda),\quad m_n > 0,\ \lambda\neq 0
\label{exampKK}
\eea
where $\lambda$ is a real arbitrary parameter 
and $\epsilon$ is the sign which appears in the boundary condition
corresponding to the Case B1b-tw (Table \re{tablethree}).
By renormalisation with
$m_n=\overline{m}_n/R$ and $\lambda =\overline{\lambda}/R$, $R$ may be
set equal to 1. 
The solutions are found at the intersections of the two curves
representing the left
and right hand sides of \re{exampKK} as functions of $\overline{m}_n$
for a given value of
$\overline{\lambda}$. This is shown for $\overline{\lambda}=2.63$ 
in Figure \re{fig1}. 
In Table \re{tabletower}, the successive masses are
given explicitly for various values of $\overline{\lambda}$. 

Let us comment these results.

For $\overline{\lambda}=\infty$, the spacing is regular
($\tan(2\pi \overline{m}_n)=0\mapsto \overline{m}_n = n/2$) which defines $n$.
The masses for $n$ even and $n$ odd belong to the towers $\epsilon=-1$
and $\epsilon=1$ respectively.
When $\overline{\lambda}$ decreases, deviations from the regularity appear. 
In fact $\overline{m}_n$ decreases continuously with $\overline{\lambda}$. 
When one reaches $\overline{\lambda}=-\infty$ 
(jumping over $\overline{\lambda}=0$ which is in principle forbidden), 
one finds again the regularity 
but with $n$ displaced by two units.

It is useful to study Eq. \re{exampKK} for very small values of $m_n$. 
The equation reduces to (up to a factor $m_n$)
\be
\overline{m}_n^2=
\overline{\lambda}\left(\overline{\lambda}+\frac{1}{\pi}\right)\ .
\label{smallm}
\end{equation} 
We see that $\overline{m}_n^2$ may become very small in two cases: 
when $\overline{\lambda}\rightarrow 0$ and when
$\overline{\lambda}\rightarrow -1/\pi$. This justifies two properties of Table \re{tabletower}. First, 
looking at the column indexed by $n=1$, one sees that $\overline{m}_1$ goes to zero when 
$\overline{\lambda}$
decreases to zero with no states 
existing anymore for $\overline{\lambda}$ negative. Then, looking at column $n=2$, 
one sees that $\overline{m}_2$ goes to zero when $\overline{\lambda}$ decreases to
$\overline{\lambda}=-1/\pi\approx -0.31831$ with again no states 
existing anymore for $\overline{\lambda}<-1/\pi$.

When $\overline{\lambda}$ is of the form of $p+1/4$ or $p+3/4$ where $p$ is an integer, 
there exists a mass in the tower 
which is exactly equal to $\overline{m}_n=|\overline{\lambda}|$ 
(for example see $\overline{\lambda}=\{0.25,-0.25,-0.75\}$ in the Table).  

\section{Conclusions}{\label{conclu}}

Coming back critically to the elementary Kaluza-Klein towers, we have studied in detail
the case of a real five dimensional scalar field assumed to propagate in the bulk.

The key determination of the KK mass squared states 
results
from the diagonalisation of the operator $P_s^2$. The physically observable $m^2$ are real 
and hence $P_s^2$ must be a symmetric operator. Analyzing mathematically this restriction, we 
have considered
all the allowed relevant boundary conditions. For many of them, $P_s$ is not symmetric. 
One should remark that the ``box'' boundary conditions ($\phi(0)=0,\ \phi(2\pi R)=0$, see
Case C5 in \re{tablebound}) 
appear in the set. Moreover the ``periodic'' or ``antiperiodic'' 
boundary conditions appear as particular cases of more general boundary conditions
(see Case C1 in Table \re{tablebound}).

All the corresponding KK states have then been
deduced from our general boundary conditions, 

Summarizing our results, we found that, in general, the KK towers consist of regular 
(equally spaced) and quite often of non regular sequences of states with in a few cases 
a zero mass state at the bottom. We found that for some specific values of the parameters
fixing the boundary conditions, some lonely KK states may exist. For completenes, we also
looked at the possiblity of tachyon KK states ($m^2<0$) and found that they would appear as
isolated states. For a massive ($M$) five dimensional scalar field, 
all the KK mass squared states are simply shifted by $M^2$. 

We have seen that our new boundary conditions for scalar fields
lead to original physical consequences. Extensions of our approach to higher 
dimensional spaces should be considered. Along the same line, a careful investigation, a priori non trivial, of
the physical conditions on boundary conditions for higher spins should be carried on.  
In particular spin one and spin two deserve special studies. 
   
In our work, essentially no dynamical consideration have been made as to the possibility 
of finding experimental evidence for the existence of the potentially predicted KK mass states.
Considering that the SO(1,4) invariance is anyhow due to be broken, a word of caution
has been put forward about the generally implicitly accepted view that there is just 
the same usual speed of light on the brane and in the bulk 
and hence a unique relation between energy and length.


\vskip 0.5 cm
\begin{table}
\caption{Table of Allowed Boundary Conditions for $P_s^2$ symmetric}
{\label{tablebound}}
\vspace{1 cm}
\hspace{4 cm}
\scriptsize{
\begin{tabular}{|l|l|}
\hline
\multicolumn{2}{|c|}
      {Boundary Conditions}\\
\multicolumn{2}{|c|}
       {$\bigl[{\rm{Symmetric}}\ (P^s)^2\bigr]$}
        \\ \hline
    Case&Boundary Conditions
      \\ \hline\hline
\multicolumn{2}{|l|}
       {Four boundary relations}
        \\ \hline
    A               &$\phi(0)=0 $
                      \\
                    &$\phi(2\pi R)=0 $
                      \\
                    &$\partial_s\phi(0)=0$
                      \\
                    &$\partial_s\phi(2\pi R)=0$
       \\ \hline
\multicolumn{2}{|l|}
       {Three boundary relations}
        \\ \hline
    B1              & $\phi(2\pi R)=\lambda_1\phi(0) $
                       \\
                    & $\partial_s\phi(0)=\lambda_2\phi(0) $
                       \\
                    &$\partial_s\phi(2\pi R)=\lambda_3\phi(0)$
       \\ \hline
    B2              & $\phi(0)=0 $
                       \\
                    & $\phi(2\pi R)=\mu_1\partial_s\phi(0) $
                       \\
                    &$\partial_s\phi(2\pi R)=\mu_2\partial_s\phi(0)$
       \\ \hline
    B3              & $\phi(0)=0 $
                       \\
                    & $\partial_s\phi(0)=0 $
                       \\
                    &$\partial_s\phi(2\pi R)=\nu\phi(2\pi R)$
       \\ \hline
    B4              & $\phi(0)=0 $
                       \\
                    &$\phi(2\pi R)=0$
                       \\
                    & $\partial_s\phi(0)=0 $
       \\ \hline
\multicolumn{2}{|l|}
       {Two boundary relations}
        \\ \hline
    C1              & $\phi(2\pi R)=\alpha_1\phi(0)
    +\alpha_2\partial_s\phi(0) $
                       \\
                    & $\partial_s\phi(2\pi R)=\alpha_3\phi(0)
                    +\alpha_4\partial_s\phi(0) $
                       \\
                    & $\alpha_1\alpha_4-\alpha_2\alpha_3=1$
       \\ \hline
    C2              & $\partial_s\phi(0)=\rho_1\phi(0) $
                       \\
                    & $\partial_s\phi(2\pi R)=\rho_2\phi(2\pi R) $
       \\ \hline
    C3              & $\phi(0)=0 $
                       \\
                    & $\partial_s\phi(2\pi R)=\kappa\phi(2\pi R) $
       \\ \hline
    C4              & $\phi(2\pi R)=0 $
                       \\
                    & $\partial_s\phi(0)=\zeta\phi(0) $
       \\ \hline
    C5              & $\phi(0)=0 $
                       \\
                    & $\phi(2\pi R)=0 $
      \\ \hline
\end{tabular}
   }
\end{table}

\newpage

\begin{table}
\caption{Table of Kaluza-Klein towers for three boundary relations}
{\label{tablethree}}
\vspace{1 cm}
\scriptsize{
\begin{tabular}{|l|l|l|}
\hline
\multicolumn{3}{|c|}
      {Boundary Conditions and Towers}\\
\multicolumn{3}{|c|}
       {for a Real Scalar Field $m_n > 0$}\\
\multicolumn{3}{|c|}
       {Three boundary relations}
        \\ \hline \hline
    Case&Boundaries\ \ $(\epsilon^2=1)$ &Type of Tower
      \\ \hline\hline
    B1a-tw &$\phi(2\pi R)=\epsilon \phi(0) $&
                    $\cos(2\pi m_n R)=\epsilon$
           $\ \left(\tau_n=\frac{m_n}{\lambda}\sigma_n\right)$
      \\
               &$\partial_s\phi(0)=\lambda \phi(0)
          \quad\quad\quad \left(\lambda\neq 0\right)$ &
                            $\mapsto m_n=\frac{n}{2R}$
      \\
               &$\partial_s\phi(2\pi R)=\epsilon\lambda \phi(0)$&
                      \ \ $\epsilon=\phantom{-}1\rightarrow n$ even
      \\
               &                 &\ \ $\epsilon=-1\rightarrow n$ odd
      \\ \hline
    B1b-tw &$\phi(2\pi R)=\epsilon \phi(0) $& $\tan(2\pi m_n R)
                      =\frac{2 m_n \lambda}{m_n^2-\lambda^2}$
            $\   \left(\tau_n=\frac{m_n}{\lambda}\sigma_n\right)$
      \\
            &$\partial_s\phi(0)=\lambda \phi(0)
            \quad\quad\quad \left(\lambda\neq 0\right)$&
                              if sign$\left(\sin(2\pi m_n R)\right)
                      =\epsilon$ sign$\left(\lambda\right)$
      \\
               &$\partial_s\phi(2\pi R)=-\epsilon\lambda \phi(0)$&\ \ $(m_n\neq 0)$
      \\ \hline
    B1c-tw &$\phi(2\pi R)=\epsilon \phi(0) $&
                    $\cos(2\pi m_n R)=\epsilon$
           $\ \left(\sigma_n=0\right)$
      \\
               &$\partial_s\phi(0)=0$ &
                            $\mapsto m_n=\frac{n}{2R}$
      \\
               &$\partial_s\phi(2\pi R)=0$&
                      \ \ $\epsilon=\phantom{-}1\rightarrow n$ even
      \\
               &                 &\ \ $\epsilon=-1\rightarrow n$ odd
      \\ \hline
    B2-tw & $\phi(0)=0 $& $\cos(2\pi m_n R)=\epsilon\ \
    \left(\tau_n=0\right)$
       \\
              &$\phi(2\pi R)=0$  &  $\mapsto m_n=\frac{n}{2R}$
       \\
              &$\partial_s\phi(2\pi R)=\epsilon\partial_s\phi(0)$  &
                     \ \ $\epsilon=\phantom{-}1\rightarrow n$ even
      \\
               &                 &\ \ $\epsilon=-1\rightarrow n$ odd
    \\ \hline
\end{tabular}
      }
\end{table}

\newpage

\begin{table}
\caption{Table of Kaluza-Klein towers for two boundary relations. In Case C1c-tw,
$F(m)$ is a solution of $(\alpha_1^2 m^2-  \ m^2  +  \alpha_3^2 )F^2
+ 2 (\alpha_1 \alpha_2 m^2 + \alpha_4 \alpha_3 )m F 
+(\alpha_2^2 m^2 + \alpha_4^2  -  1)m^2=0$ and 
$G(m,F)=\frac
{- \alpha_3 F^2 + (\alpha_1  - \alpha_4 )m F + \alpha_2 m^2 } 
{\alpha_1 m F^2 + (\alpha_2 m^2 + \alpha_3 ) F + \alpha_4 m }$}
{\label{tabletwo}}
\vspace{1 cm}
\scriptsize{
\begin{tabular}{|l|l|l|}
\hline
\multicolumn{3}{|c|}
      {Boundary Conditions and Towers}\\
\multicolumn{3}{|c|}
       {for a Real Scalar Field $m_n > 0$}\\
\multicolumn{3}{|c|}
       {Two boundary relations}
        \\ \hline \hline
    Case&Boundaries\ \ $(\epsilon^2=1)$ &Type of Tower
      \\ \hline\hline
   C1a-tw &  $\phi(2\pi R)=\epsilon \phi(0)$& $\cos(2\pi m_n R)=
   \epsilon$
                   $\ \ \left(\tau_n=0\right)$
      \\
               & $\partial_s\phi(2\pi R)=\alpha_3 \phi(0)+\epsilon
               \partial_s\phi(0)$
                         &$ \mapsto m_n=\frac{n}{2R}$
        \\
                &     &    \ \ $\epsilon=\phantom{-}1\rightarrow n$
                even
      \\
               &                 &\ \ $\epsilon=-1\rightarrow n$ odd
      \\ \hline
   C1b-tw &  $\phi(2\pi R)=\epsilon \phi(0)
   +\alpha_2\partial_s\phi(0)$&
              $\cos(2\pi m_n R)=\epsilon\ \ \left(\sigma_n=0\right)$
     \\
               & $\partial_s\phi(2\pi R)=\epsilon \partial_s\phi(0)$
                         & $ \mapsto m_n=\frac{n}{2R}$
        \\
                &     &    \ \ $\epsilon=\phantom{-}1\rightarrow n$
                even
      \\
               &                 &\ \ $\epsilon=-1\rightarrow n$ odd
      \\ \hline
   C1c-tw & $\phi(2\pi R)=\alpha_1\phi(0)
   +\alpha_2\partial_s\phi(0)$ &
                                         $\tan(2\pi m_n R)=G(m_n,F(m_n))$
     \\
              &$\partial_s\phi(2\pi R)=\alpha_3\phi(0)
              +\alpha_4\partial_s\phi(0)$
                                      & $\tau_n/\sigma_n=F(m_n)$
       \\
              &$\alpha_1\alpha_4-\alpha_3\alpha_2=1$ &see caption
      \\ \hline
   C2a-tw &  $\partial_s\phi(0)=\rho\phi(0)\quad\quad\quad
   \left(\rho\neq 0\right)$
        & $\sin(2\pi m_n R)=0$ $\ \ \left(\tau_n=\frac{m_n}{\rho}
        \sigma_n\right)$
             \\
            &  $\partial_s\phi(2\pi R)=\rho\phi(2\pi R)$ & $\mapsto
            m_n=\frac{n}{2R}$
              \\
            &                                            & $n$ integer
   \\ \hline
   C2b-tw &  $\partial_s\phi(0)=0$
        & $\sin(2\pi m_n R)=0$ $\ \ \left(\sigma_n=0\right)$
             \\
            &  $\partial_s\phi(2\pi R)=0$ & $\mapsto m_n=\frac{n}{2R}$
              \\
            &                                            & $n$ integer
   \\ \hline
   C2c-tw &  $\partial_s\phi(0)=\rho_1\phi(0)$
                       \quad\quad\quad$\left( \rho_1\neq 0\right)$ &
                   $\tan(2\pi m_n R)=\frac{m_n(\rho_1-\rho_2)}
                   {m_n^2+\rho_1\rho_2}$
                        $\ \ \left(\tau_n=\frac{m_n}{\rho_1}
                        \sigma_n\right)$
       \\
            &  $\partial_s\phi(2\pi R)=\rho_2\phi(2\pi R)\quad\left(
            \rho_2\neq \rho_1\right)$&
   \\ \hline
   C2d-tw &  $\partial_s\phi(0)=0$  &
                   $\cot(2\pi m_n R)=-\frac{m_n}{\rho_2}$
                        $\ \ \left(\sigma_n=0\right)$
       \\
            &  $\partial_s\phi(2\pi R)=\rho_2\phi(2\pi R)\quad\left(
            \rho_2\neq 0\right)$&
   \\ \hline
    C3a-tw &  $\phi(0)=0$  &   $\tan(2\pi m_n R)=\frac{m_n}{\kappa}
    $$\ \ \left(\tau_n=0\right)$
        \\
              &  $\partial_s\phi(2\pi R)=\kappa\phi(2\pi R)
              \quad\left(\kappa\neq 0\right)$&
      \\ \hline
    C3b-tw &  $\phi(0)=0$  & $\cos(2\pi m_n R)=0$  $\ \
    \left(\tau_n=0\right)$
        \\
              &  $\partial_s\phi(2\pi R)=0$& $\mapsto m_n=\frac{n}{2R}
              +\frac{1}{4R}$
               \\
            &                                            &         $n$
            integer
      \\ \hline
    C4a-tw &  $\phi(2\pi R)=0$ &   $\tan(2\pi m_n R)=-\frac{m_n}{\lambda}
    $
        $\ \left(\tau_n=\frac{m_n}{\lambda}\sigma_n\right)$
    \\
             & $\partial_s\phi(0)=\lambda\phi(0)
             \quad\quad\left(\lambda\neq 0\right)$ &
     \\ \hline
    C4b-tw &  $\phi(2\pi R)=0$ &   $\cos(2\pi m_n R)=0$
        $\ \left(\sigma_n=0\right)$
    \\
             & $\partial_s\phi(0)=0$ &$\mapsto m_n=\frac{n}{2R}
             +\frac{1}{4R}$
               \\
          &                                            &         $n$
          integer
     \\ \hline
    C5-tw & $\phi(0)=0$ &   $\sin(2\pi m_n R)=0 \ \ \left(\tau_n=0\right)
    $     \\
             & $\phi(2\pi R)=0$ & $\mapsto m_n=\frac{n}{2R}$
               \\
            &                                            &         $n$
            integer
      \\ \hline
\end{tabular}
   }
\end{table}

\newpage

\begin{table}
\caption{Table of Kaluza-Klein towers with a zero mass state}
{\label{tablezero}}
\vspace{1 cm}
\scriptsize{
\begin{tabular}{|l|l|l|}
\hline
\multicolumn{3}{|c|}
      {Real Scalar Tower with a natural zero mass $m_0=0$}
        \\
\multicolumn{3}{|c|}
      {at the bottom}
        \\ \hline \hline
    Case&Boundaries &Tower
      \\ \hline\hline
    B1-z &$\phi(2\pi R)=\phi(0) $&
                    $\cos(2\pi m_n R)=1$
           $\ \left(\sigma_n=0\right)$
      \\
               &$\partial_s\phi(0)=0$&
                            $\mapsto m_n=\frac{n}{2R}$
      \\
               &$\partial_s\phi(2\pi R)=0$&
                      \ \ $n$ even
          \\ \hline
   C1-z &  $\phi(2\pi R)= \phi(0)+\alpha_2\partial_s\phi(0)$&
              $\cos(2\pi m_n R)=1\ \ \left(\sigma_n=0\right)$
     \\
               & $\partial_s\phi(2\pi R)= \partial_s\phi(0)$
                         & $ \mapsto m_n=\frac{n}{2R}$
        \\
                &     &   $n$ even
      \\ \hline
   C2-z &  $\partial_s\phi(0)=0$
        & $\sin(2\pi m_n R)=0$ $\ \ \left(\sigma_n=0\right)$
             \\
            &  $\partial_s\phi(2\pi R)=0$ & $\mapsto m_n=\frac{n}{2R}$
              \\
            &                                            & $n$ integer
   \\ \hline
\end{tabular}
       }
\end{table}

\newpage

\begin{table}
\caption{Table of Lonely Mass Cases}
{\label{tablespecial}}
\vspace{1 cm}
\scriptsize{
\begin{tabular}{|l|l|l|}
\hline
\multicolumn{3}{|c|}
      {Boundary Conditions, Lonely Mass Cases}\\
\multicolumn{3}{|c|}
       {for a Real Scalar Field $m\neq 0$}
        \\ \hline \hline
    Case&Boundaries &Type of Tower
      \\ \hline\hline
    B1-s &$\phi(2\pi R)=\lambda_1 \phi(0)$ &
                   $m^2=\frac{\lambda_1^2-\lambda_3^2}{\lambda_2^2-1}$
                 $\ \ \left(\sigma=\frac{\lambda_1}{n}\tau\right)$
      \\
             &$\partial_s\phi(0)=\lambda_2 \phi(0)$$\ \ \ \
             \left(\lambda_2^2\neq 1\right)$
   &if $\tan(2\pi m R)=\frac{m(\lambda_1\lambda_2-\lambda_3)}
   {\lambda_1\lambda_3+\lambda_2 m^2}$
      \\
               &$\partial_s\phi(2\pi R)=\lambda_3 \phi(0)$&
      \\ \hline
    B2-s &$\phi(0)=0 $& $m^2=\frac{1-\mu_2^2}{\mu_1^2}$
                $\ \ \left(\tau=0\right)$
      \\
               &$\phi(2\pi R)=\mu_1 \partial_s\phi(0)$$\ \ \ \
               \left(\mu_1\neq 0\right)$&
     if $\tan(2\pi m R)=\frac{m\mu_1}{\mu_2}$
     \\
               &$\partial_s\phi(2\pi R)=\mu_2 \partial_s\phi(0)
               $&
        \\ \hline
   C1a-s & $\phi(2\pi R)=\alpha_1\phi(0)
   +\alpha_2\partial_s\phi(0)$ &
                      $m^2=\frac{1-\alpha_4^2}{\alpha_2^2}$
                  $\ \ \left(\tau=0\right)$
        \\
              &$\partial_s\phi(2\pi R)=\alpha_3\phi(0)
              +\alpha_4\partial_s\phi(0)$
                         &if $\tan(2\pi m R)=\frac{m\alpha_2}
                         {\alpha_4}$
       \\
              &$\alpha_1\alpha_4-\alpha_3\alpha_2=1$$\ \ \ \
              \left(\alpha_2\neq 0\right)$
        &
      \\ \hline
   C1b-s & $\phi(2\pi R)=\alpha_1\phi(0)
   +\alpha_2\partial_s\phi(0)$ &
                      $m^2=\frac{\alpha_3^2}{1-\alpha_1^2}$
                  $\ \ \left(\sigma=0\right)$
       \\
              &$\partial_s\phi(2\pi R)=\alpha_3\phi(0)
              +\alpha_4\partial_s\phi(0)$
                         &if $\tan(2\pi m R)=-\frac{\alpha_3}{\alpha_1
                         m}$
        \\
              &$\alpha_1\alpha_4-\alpha_3\alpha_4=1$$\ \ \
              \left(\alpha_1^2\neq 1\right)$&
      \\ \hline
   C2-s &  $\partial_s\phi(0)=\rho_1\phi(0)$ &
   $m^2=-\rho_1\rho_2$
                  $\ \ \left(\sigma=\frac{\rho_1}{m}\tau\right)$
       \\
            &  $\partial_s\phi(2\pi R)=\rho_2\phi(2\pi R)$ &if
            $\sqrt{-\rho_1\rho_2}=\left(\frac{n}{2}+\frac{1}{4}\right)/R$
       \\
            &  $\rho_2\neq \rho_1$, $\rho_1\rho_2<0$   &            
 \\ \hline
\end{tabular}
      }
\end{table}

\newpage
\begin{table}
\caption{Table of tachyons. Three boundary relations}
{\label{tabletachyons1}}
\vspace{1 cm}
\scriptsize{
\begin{tabular}{|l|l|l|}
\hline
\multicolumn{3}{|c|}
      {Tachyons}\\
\multicolumn{3}{|c|}
       {for a Real Scalar Field $m^2 = - h^2<0$}
        \\
\multicolumn{3}{|c|}
      {Three boundary relations}
     \\\hline \hline
    Case&Boundaries\ \ $(\epsilon^2=1)$ &Type of Tower
      \\ \hline\hline
    B1a-t &$\phi(2\pi R)=\lambda_2 \phi(0) $&
                    $h=\lambda_1$
           $\ \left(\widetilde{\tau}= 0\right)$
      \\
               &$\partial_s\phi(0)=\lambda_1 \phi(0)$&
      \\
               &$\partial_s\phi(2\pi R)=\lambda_1\lambda_2 \phi(0)$&
      \\
               &$\quad\left( e^{2\pi \lambda_1 R}=\lambda_2\right)$ &
      \\ \hline
    B1b-t &$\phi(2\pi R)=\lambda_2 \phi(0) $&
                    $h=-\lambda_1$
           $\ \left(\widetilde{\sigma}= 0\right)$
      \\
               &$\partial_s\phi(0)=\lambda_1 \phi(0)$&
      \\
               &$\partial_s\phi(2\pi R)=\lambda_1\lambda_2 \phi(0)$&
      \\
               &$\quad\left( e^{-2\pi \lambda_1 R}=\lambda_2\right)$ &
      \\ \hline
    B1c-t &$\phi(2\pi R)=\lambda_2 \phi(0) $&
                    $h=\epsilon x$
           $\ \left(\widetilde{\sigma}= \frac{\epsilon x+\lambda_1}
                                             {\epsilon x-\lambda_1}\
                                             \widetilde{\tau}\right)$
      \\
               &$\partial_s\phi(0)=\lambda_1 \phi(0)$&
      \\
               &$\partial_s\phi(2\pi R)=\lambda_3 \phi(0)$&
      \\
               &$\quad\left(x=\sqrt{\frac{\lambda_3^2-\lambda_1^2}
               {\lambda_2^2-1}}>0,\
               e^{2\pi \epsilon x R}=\frac{\epsilon x-\lambda_1}
                             {\epsilon\lambda_2x-\lambda_3}\right)$ &
      \\ \hline
    B2-t        & $\phi(0)=0 $& $h=\epsilon x \
    \left(\widetilde{\sigma}=-\widetilde{\tau}\right)$
       \\
              &$\phi(2\pi R)=\mu_1\partial_s\phi(0)$  &
       \\
              &$\partial_s\phi(2\pi R)=\mu_2\partial_s\phi(0)$  &
      \\
               & $\quad \left(x=\frac{\sqrt{\mu_2^2-1}}{\mu_1}>0,\
                e^{2\pi \epsilon x R}=\frac{1}
                                          {\mu_2-x\mu_1}
                                          \right)$       &
    \\ \hline
\end{tabular}
     }
\end{table}

\newpage

\begin{table}
\caption{Table of tachyons. Two boundary relations }
{\label{tabletachyons2}}
\vspace{0.5 cm}
    \tiny{
\begin{tabular}{|l|l|l|}
\hline
\multicolumn{3}{|c|}
      {Tachyons for a Real Scalar Field $m^2 = - h^2<0$}
        \\
\multicolumn{3}{|c|}
      {Two boundary relations}
     \\\hline \hline
    Case&Boundaries\ \ $(\epsilon^2=1)$\ \ &Type of Tower
      \\ \hline\hline
   C1a-t & $\phi(2\pi R)=\alpha_1\phi(0)
   +\alpha_2\partial_s\phi(0)
                           \ \ \left(\alpha_2\neq 0\right)$ &
                                        $h=\epsilon x$ \
               \\
              &$\partial_s\phi(2\pi R)=\alpha_3\phi(0)
              +\alpha_1\partial_s\phi(0)$
                                      &
       \\
              &\quad$\left(\alpha_1^2-\alpha_3\alpha_2=1\right)$ &
       \\
              &$\quad\left(x=\frac{\sqrt{\alpha_1^2-1}}{\alpha_2}>0,\
                e^{2\pi \epsilon x R}= \alpha_1+\epsilon x\alpha_2
                \right)$               &
      \\ \hline
   C1b-t & $\phi(2\pi R)=\alpha_1\phi(0)
   +\alpha_2\partial_s\phi(0)
                                \ \ \left(\alpha_2\neq 0\right)$&
                                        $h=\epsilon x$
     \\
              &$\partial_s\phi(2\pi R)=\alpha_3\phi(0)
              +\alpha_4\partial_s\phi(0)$
                                      & $\left(\widetilde{\tau}=
             \left(1-2\epsilon\alpha_1\alpha_2-2\alpha_1^2\right)
             \widetilde{\sigma}\right)$
       \\
              &\quad$\left(\alpha_1\alpha_4-\alpha_3\alpha_2=1,\
              \alpha_4\neq\alpha_1\right)$  &
       \\
              &$\quad\left(x=\frac{\sqrt{\alpha_1^2-1}}{\alpha_2}>0,\
                e^{2\pi \epsilon x R}= \alpha_1+\epsilon x\alpha_2
                \right)$               &
      \\ \hline
   C1ct & $\phi(2\pi R)=\alpha_1\phi(0)
   +\alpha_2\partial_s\phi(0)
                               \ \ \left(\alpha_2\neq 0\right)$ &
                                        $h=x\ \left(\widetilde{\tau}=0
                                        \right)$
     \\
              &$\partial_s\phi(2\pi R)=\alpha_3\phi(0)
              +\alpha_4\partial_s\phi(0)$
                                      &
       \\
              &\quad$\left(\alpha_1\alpha_4-\alpha_3\alpha_2=1,
               \ \left(\alpha_1+\alpha_4\right)^2>4\right)$  &
       \\
              &$\quad\left( x\right.$ real root of&
       \\
               &$\quad\left.\left[\alpha_2^2 x^2+(\alpha_1-\alpha_4)
               \alpha_2x
                               +(1-\alpha_1\alpha_4)=0\right]\right)$&
       \\
               &\quad$\left(e^{2\pi x R}= \alpha_1+x\alpha_2 \right)$
               &
      \\ \hline
   C1d-t & $\phi(2\pi R)=\alpha_1\phi(0)
   +\alpha_2\partial_s\phi(0)
                                \ \ \left(\alpha_2\neq 0\right)$ &
                                        $h=$ real root of
     \\
              &$\partial_s\phi(2\pi R)=\alpha_3\phi(0)
              +\alpha_4\partial_s\phi(0)$
                                      & $\left[\left(
                          \alpha_2^2h^2-\alpha_2\left(\alpha_1+\alpha_
                          4\right)h
                                    +\left(\alpha_1\alpha_4-1\right)
                                              \right)e^{4\pi h R}
                                        \right.$
       \\
              &\quad$\left(\alpha_1\alpha_4-\alpha_3\alpha_2=1
                           \right)$
                                      &$\quad\quad+ 4\alpha_2 he^{2\pi
                                      h R}$
       \\
              &                   & $-\left.
                          \left(

                                   \alpha_2^2h^2+\alpha_2\left(\alpha_
                                   1+\alpha_4\right)h
                                    +\left(\alpha_1\alpha_4-1\right)
                          \right)=0
                                        \right]$
       \\
               &              & $\widetilde{\sigma}=
                          \frac{\left(\alpha_2 h-\alpha_1\right)
                          e^{2\pi h R}+1}
                               {\left(\alpha_1+\alpha_2 h - e^{2\pi h
                               R}\right)e^{2\pi h R}}
                                \ \widetilde{\tau}$
      \\ \hline
   C1e-t & $\phi(2\pi R)=\alpha_1\phi(0)$ &
                                        
     \\
              &$\partial_s\phi(2\pi R)=\alpha_3\phi(0)+\frac{1}
              {\alpha_1}\partial_s\phi(0)$
                                      &
                                $h=\frac{\alpha_1\alpha_3}{\alpha_1^2-1}
                                     \ \left(\widetilde{\tau}=0
                                     \right)$
       \\                   &  $\quad\left(e^{2\pi h R}=
       \alpha_1\right)$
                      &
      \\ \hline
   C1f-t & $\phi(2\pi R)=\alpha_1\phi(0)$ &
                                        $h=$ real root of
     \\
              &$\partial_s\phi(2\pi R)=\alpha_3\phi(0)
              +\frac{1}{\alpha_1}\partial_s\phi(0)$
                                      & $\left[
                              \left(
                             \alpha_1\alpha_3-\left(\alpha_1^2+1\right)h
                              \right)e^{4\pi h R}
                                        \right.$
       \\
               &  $\quad\left(e^{2\pi h R}=\alpha_1\right)$
                                &  $\quad\quad + 4\alpha_1h e^{2\pi
                                hR}$
      \\
              &                 & $\left.
                                       -
                                       \left(\alpha_1\alpha_3
                                 +\left(\alpha_1^2+1\right)h\right)=0
                                  \right]$
      \\
              &            & $\widetilde{\sigma}=
                               \frac{1-\alpha_1 e^{2\pi h R}}
                                    {\left(\alpha_1-e^{2\pi h R}
                                    \right)e^{2\pi h R}}
                             \ \widetilde{\tau}$
        \\ \hline
   C2a-t &  $\partial_s\phi(0)=\rho\phi(0)\quad\quad\quad
   \left(\rho\neq 0\right)$
        & $h=\rho\ \left(\widetilde{\rho}=0\right)$
             \\
            &  $\partial_s\phi(2\pi R)=\rho\phi(2\pi R)$ &
   \\ \hline
   C2b-t &  $\partial_s\phi(0)=\rho\phi(0)\quad\quad\quad
   \left(\rho\neq 0\right)$
        & $h=-\rho\ \left(\widetilde{\sigma}=0\right)$
             \\
            &  $\partial_s\phi(2\pi R)=\rho\phi(2\pi R)$ &
   \\ \hline
   C2c-t &  $\partial_s\phi(0)=\rho_1\phi(0)$    &
                        $h$= real root of
       \\
            &  $\partial_s\phi(2\pi R)=\rho_2\phi(2\pi R)$&
                       $\left[e^{4\pi h R}=
                        \frac{h^2-\left(\rho_1-\rho_2\right)
                        h-\rho_1\rho_2}
                             {h^2 + \left(\rho_1-\rho_2\right)
                             h-\rho_1\rho_2}
                        \right],\  \left(h^2\neq \rho_1^2\right)$
     \\
             &          &$\left(\widetilde{\sigma}=\frac{h+\rho_1}
                                                      {h-\rho_1}
                              \ \widetilde{\tau}
                          \right)$
   \\ \hline
    C3-t        &  $\phi(0)=0$  & $h$= real root of
        \\
              &  $\partial_s\phi(2\pi R)=\kappa\phi(2\pi R)\
              \left(\kappa\neq 0\right)$
                        &$\left[e^{4\pi h R}=
                        \frac{\kappa +h}
                             {\kappa - h}
                        \right],\  \left(h^2\neq \kappa^2\right)$
     \\
             &          &$\left(\widetilde{\sigma}=-
                              \widetilde{\tau}
                          \right)$
      \\ \hline
    C4-t &  $\phi(2\pi R)=0$ &    $h$= real root of
    \\
             & $\partial_s\phi(0)=\lambda\phi(0)
             \quad\quad\left(\lambda\neq 0\right)$ &
                  $\left[e^{4\pi h R}=
                        \frac{\lambda - h}
                             {\lambda + h}
                        \right],\  \left(h^2\neq \lambda^2\right)$
     \\
             &          &$\left(\widetilde{\sigma}=\frac{h+\lambda}
                                                      {h-\lambda}
                              \ \widetilde{\tau}
                          \right)$
      \\ \hline
\end{tabular}
    }
\end{table}

\newpage

\begin{table}
\caption{Case B1b-tw. The $\overline{m}_n$-towers as a function of
$\overline{\lambda}$ }
\label{tabletower}
\vspace{1 cm}
\hspace*{-2 cm}
\scriptsize{
\begin{tabular}{|c|l|c|c|c|c|c|c|c|c|c|c|c|}
\hline
   \multicolumn{2}{|c|}{}        &\multicolumn{11}{c|}{$n$}
        \\
\cline{3-13}
    \multicolumn{2}{|c|}{} &1&2&3&4&5&6&7&8&9&10&11
          \\
\cline{3-13}
   \multicolumn{2}{|c|}{}        &\multicolumn{11}{c|}{$\epsilon$}
        \\
\cline{3-13}
    \multicolumn{2}{|c|}{} &$+1$&$-1$
                           &$+1$&$-1$&$+1$
                           &$-1$&$+1$&$-1$
                           &$+1$&$-1$&$+1$
          \\
\hline
 &$\phantom{-\,}$$\infty$
 &0.5   &1.   &1.5    &2.    &2.5   &3.    &3.5   &4.
 &4.5   &5   &5.5
         \\
\cline{2-13}
 &$\phantom{-\,}$100.
 &0.499 &0.997 &1.496 &1.994 &2.493 &2.991 &3.489 &3.988
 &4.486 &4.985 &5.483
         \\
\cline{2-13}
 &$\phantom{-\,}$10.
 &0.485 &0.97  &1.455 &1.94  &2.425 &2.91  &3.396 &3.883
 &4.369 &4.857 &5.344
         \\
\cline{2-13}
 &$\phantom{-\,}$1.
 &0.384 &0.788 &1.219 &1.672 &2.14  &2.617 &3.1   &3.587
 &4.077 &4.569 &5.063
         \\
\cline{2-13}
 &$\phantom{-\,}$0.25
&0.25   &0.622 &1.073 &1.551 &2.039 &2.532 &3.027 &3.523
&4.02   &4.518 &5.016
         \\
\cline{2-13}
 &$\phantom{-\,}$0.1
 &0.17  &0.557 &1.031 &1.521 &2.016 &2.513 &3.011 &3.51
 &4.008 &4.508 &5.007
         \\
\cline{2-13}
 &$\phantom{-\,}$0.01
 &0.057 &0.507 &1.004 &1.503 &2.002 &2.502 &3.002 &3.501 &4.001
 &4.501 &5.001
         \\
\cline{2-13}
${\overline{\lambda}}$
 & $\phantom{-\,}$0.001
 &0.018 &0.5 &1. &1.5 &2. &2.5 &3. &3.5 &4.
 &4.501 &5.001
         \\
\cline{2-13}
 & $-\,$0.001
 &      &0.5   &1.     &1.5   &2.     &2.5   &3.     &3.5
 &4.     &4.5   &5.
        \\
\cline{2-13}
 & $-\,$0.01
 &      &0.494 &0.997 &1.498 &1.999 &2.499 &2.999 &3.5
 &4     &4.5   &5
         \\
\cline{2-13}
 & $-\,$0.1
 &      &0.427 &0.968 &1.479 &1.984 &2.488 &2.99  &3.491 &3.993
 &4.493 &4.994
         \\
\cline{2-13}
 & $-\,$0.25
 &      &0.25  &0.916 &1.446 &1.96  &2.468 &2.974 &3.478 &3.981
 &4.483 &4.985
         \\
\cline{2-13}
 & $-\,$0.3183
&      &0.004  &0.891 &1.431 &1.949 &2.46  &2.966 &3.471 &3.975 
&4.478  &4.98
         \\  
\cline{2-13}
 & $-\,$0.3184
&      &      &0.891 &1.431 &1.949 &2.46  &2.966 &3.471 &3.975 
&4.478  &4.98
         \\  
\cline{2-13}
 & $-\,$0.75
&      &      &0.75  &1.338 &1.88  &2.404 &2.92  &3.432 &3.941
&4.447  &4.953
         \\
\cline{2-13}
 & $-\,$1.
 &      &      &0.693 &1.291 &1.842 &2.374 &2.895 &3.41  &3.921
 &4.43  &4.937
         \\
\cline{2-13}
 & $-\,$5.
 &      &      &0.534 &1.067 &1.599 &2.129 &2.656 &3.181 &3.703
 &4.224 &4.742
         \\
\cline{2-13}
 & $-\,$20.
 &      &      &0.509 &1.017 &1.525 &2.033 &2.541 &3.049 &3.557
 &4.064 &4.572
         \\
\cline{2-13}
 & $-\,$100.
 &      &      &0.502 &1.004 &1.505 &2.007 &2.508 &3.01  &3.512
 &4.013 &4.515
         \\
\cline{2-13}
 & $-\,$$\infty$
 &      &      &0.5   &1.   &1.5    &2.    &2.5   &3.    &3.5   &4.
 &4.5
         \\
\hline
\end{tabular}
     }
\end{table}

\newpage

\begin{figure}[ht]\label{fig1}
\caption{The KK tower masses at the intersection of
$\tan(2\pi \overline{m}_n)$ and 
$2\overline{\lambda}\overline{m}_n/\left(\overline{m}_n^2-\overline{\lambda}^2\right)$
for $\overline{\lambda}=2.63$, as functions of $\overline{m}_n$
(Case B1b-tw in Table \re{tablethree})}
\vskip 2 cm
\begin{center}
\epsfxsize=12cm
\epsffile{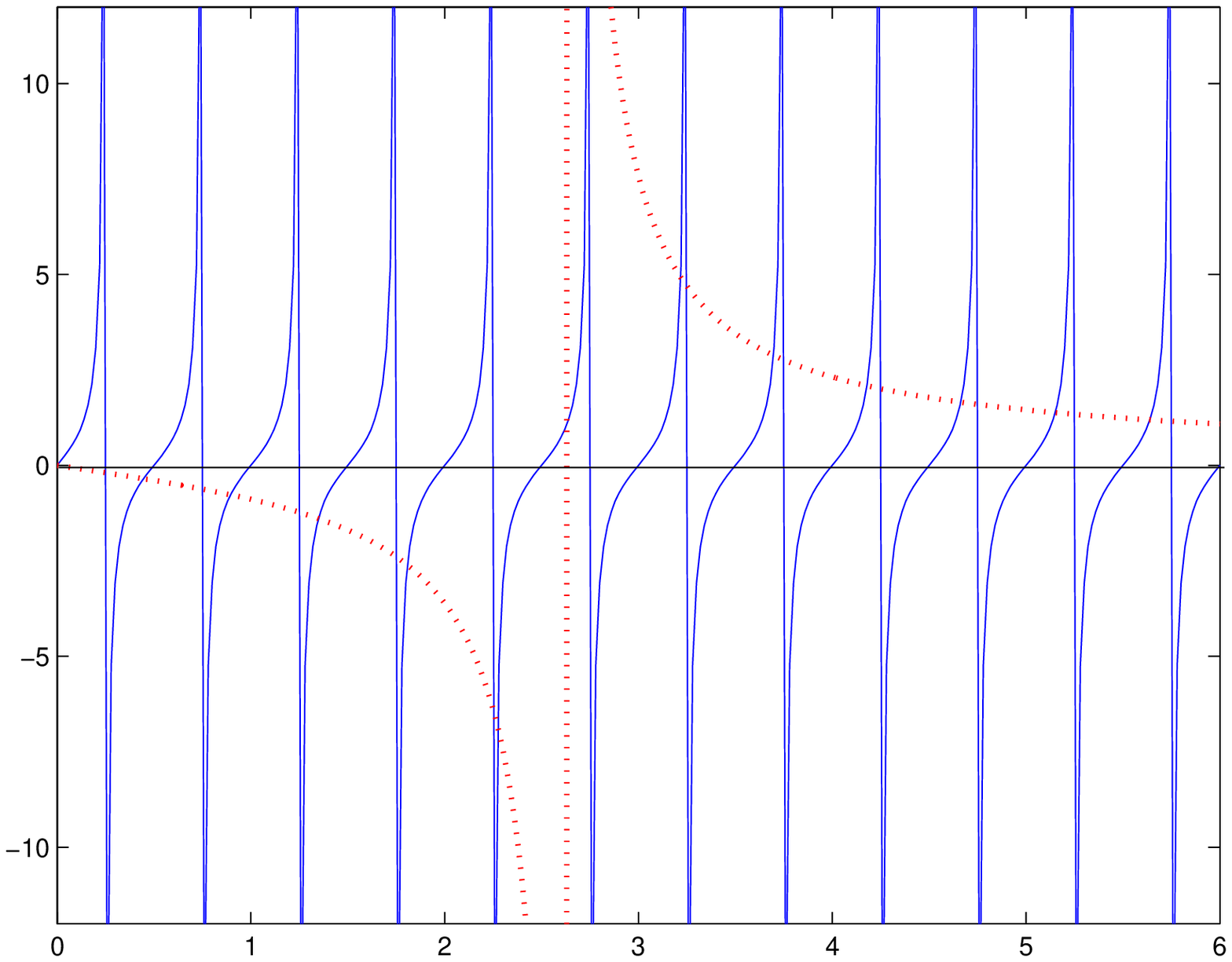}
\end{center}
\end{figure}

\goodbreak

\end{document}